\documentclass[twocolumn,aps,pra,reprint,superscriptaddress]{revtex4-1}

\usepackage{amsmath}

\usepackage{graphicx}

\usepackage[lofdepth,lotdepth, caption=false]{subfig}

\usepackage{verbatim}

\usepackage{color}

\usepackage{hyperref}

\usepackage{dcolumn}

\usepackage{bm}

\usepackage{miller}

\usepackage{color}

\usepackage{float}

\begin{document}

\title{Emergence of topologically protected states in MoTe$_{2}$ Weyl semimetal with layer stacking order}

\author{John A.~Schneeloch}

\affiliation{Department of Physics, University of Virginia, Charlottesville, Virginia 22904, USA}
\author{Chunruo Duan}
\affiliation{Department of Physics, University of Virginia, Charlottesville, Virginia 22904, USA}
\author{Junjie Yang}
\altaffiliation[Present address: ]{Department of Physics, Central Michigan University, Mount Pleasant, Michigan 44859}
\affiliation{Department of Physics, University of Virginia, Charlottesville, Virginia 22904, USA}
\author{Jun Liu}
\affiliation{Division of Materials Science \& Engineering, Ames Laboratory, Ames, Iowa 50011, USA}
\author{Xiaoping Wang}
\affiliation{Chemical and Engineering Materials Division, Oak Ridge National Laboratory, Oak Ridge, Tennessee 37831, USA}
\author{Despina Louca}
\thanks{Corresponding author}
\email{louca@virginia.edu}
\affiliation{Department of Physics, University of Virginia, Charlottesville, Virginia 22904, USA}

\begin{abstract}
Electronic tunability in crystals with weakly-bound layers can be achieved through layer stacking order. One such example is MoTe$_2$, where the low-temperature orthorhombic T$_d$ phase is topological and host to Weyl quasiparticles. The transition mechanism to the non-trivial topology is elucidated by single crystal neutron diffraction. Upon cooling from the monoclinic 1T$\prime$ to the T$_d$ phase, diffuse scattering accompanies the transition, arising from random, in-plane layer displacements, and dissipates upon entering the T$_d$ phase. Diffuse scattering is observed only in the H0L plane due to irreversible layer shifts along the c-axis that break the centrosymmetry of the monoclinic lattice.

\end{abstract}
\maketitle

Transition metal dichalcogenides (TMDs) are hosts to exotic quantum states, with electronic features that are suitable for optoelectronic and quantum technologies \cite{cite1, cite2}. Their crystal structures consist of Van der Waals-bound layers, where a change in the layer stacking can result in new properties such as superconductivity, recently observed in bilayer graphene with a \textquotedblleft magic\textquotedblright twist angle \cite{cite3}, or transition to the Weyl semimetal state \cite{cite4,cite5,cite6,cite7} reported in the T$_d$ phase of MoTe$_2$ and in the Kooi phase of Ge$_{2}$Sb$_{2}$Te$_{5}$\cite{cite8}. MoTe$_{2}$ is a prototype for understanding how stacking variations in layered materials can lead to exotic states of matter. Its crystal structure can be tuned by temperature and pressure between two phases; the 1T$\prime$, a topologically trivial phase, and the non-centrosymmetric T$_d$ phase, the host of Weyl quasiparticles. The crystal symmetry is thus essential to predicting the emergence of topologically protected states.

The mechanism of the structural transition has been of particular interest in MoTe$_2$, since Weyl quasiparticles are predicted in the low temperature phase of the non-centrosymmetric orthorhombic T$_d$ phase, protected by crystal symmetry. Early X-ray diffraction and Raman scattering measurements suggested that the high temperature 1T$\prime$ monoclinic structure belongs to the P2$_1$/m space group that preserves inversion symmetry \cite{cite23,cite24,cite25,cite26}. More recent Raman and second harmonic generation measurements indicated that the inversion symmetry of the 1T$\prime$ phase is most likely broken in thin films \cite{cite27}. 

How the stacking pattern and disorder arise in MoTe$_2$ has implications on many other Van der Waals-bound layered materials where stacking can be controlled reproducibly by temperature or pressure. Examples include transitions with pressure in WTe$_2$ \cite{cite9}, ReS$_2$ \cite{cite10}, ReSe$_2$ \cite{cite11}, MoS$_2$ \cite{cite12,cite13,cite14}, and Ta$_2$NiSe$_5$ \cite{cite15}; with temperature in RuCl$_3$ \cite{cite16}, CrX$_3$ (X=Cl, Br, I) \cite{cite17}, and CdPS$_3$ \cite{cite18}; and with either temperature or pressure in In$_2$Se$_3$ \cite{cite19,cite20} and MoTe$_2$ \cite{cite21}. In MoTe$_2$, the stability of the 1T$\prime$ and T$_d$ phases was explained through density functional theory (DFT) calculations \cite{cite22}, but neglected to investigate how the transition proceeds.

The difference between 1T$\prime$ and T$_d$ states can be illustrated by considering the nature of the layer stacking. Shown in Figs 1(a) and (d) are the crystal structures of the T$_d$ and 1T$\prime$, respectively. The T$_d$ phase can be thought of as having \textquotedblleft AA\textquotedblright layer order, where \textquotedblleft A\textquotedblright denotes an operation mapping one layer to the next. For T$_d$, this operation involves translation along the c-axis by 0.5 lattice constants and reflections about the a- and b-directions. Though there are two layers per unit cell, the operation is the same for both layers. In contrast, the layer stacking of 1T$\prime$ can be described by \textquotedblleft AB\textquotedblright, where \textquotedblleft B\textquotedblright\ denotes an operation just like \textquotedblleft A\textquotedblright\ but with an additional shift of about $\pm0.15$ lattice constants along the a-direction, where the sign depends on weather the \textquotedblleft B\textquotedblright\ is in an even or odd position in the layer sequence. Though this description of 1T$\prime$ is approximate and neglects additional intralayer distortions relative to the layers in T$_d$, it captures the binary choice of layer placement at each inter-layer boundary. The additional shifts in 1T$\prime$ result in a tilting of the unit cell with a monoclinic angle $\beta\approx93.9^{\circ}$. To elucidate the nature of the transition mechanism across the phase boundary, we employed high-resolution single crystal neutron diffraction. In this letter, we show that the diffuse scattering that appears on cooling through the 1T$\prime$ to T$_d$ transition is consistent with random layer shifts, driving the transition from ABAB layer stacking in the 1T$\prime$ phase to AAAA layer stacking in the T$_d$ phase. 

The single crystal of MoTe$_2$ was grown in a 1:25 molar ratio of Mo:Te using high purity elements (99.9999\% for both). The elements were heated together in an evacuated silica ampoule to 1050 $^{\circ}$C, held there for 24 hours, then cooled to 900 $^{\circ}$C at a rate of 0.5 $^{\circ}$C/h and quenched in liquid nitrogen. The neutron scattering measurements were performed on the single crystal diffractometer TOPAZ at Oak Ridge National Laboratory. By indexing the Bragg peaks, an orientation matrix (a.k.a. UB matrix) is defined for the data at each temperature. The lattice constants are obtained from the UB matrices. For the 295 and 260 K data, the UB matrices are defined using only the reflections from one of the twin domains. The structure factors of each Bragg reflections are calculated from the reduced and normalized data in order to determine the crystal structure. To accommodate the L-direction elongated peak shape, the intensities of the reflections are taken as integrals over a box of size $0.3 \times 0.3 \times 0.5$ in reciprocal units subtracting the average background in a shell of 0.08 reciprocal unit thickness. Atomic coordinates in this paper are defined in the convention typically used for the 1T$\prime$ phase, i.e., $b<a<c$. The lattice parameters are listed in Table 1. Data in the form of intensity as a function of momentum transfer in three-dimensional reciprocal space were collected successively at 295, 260, 240, and 100 K, and averaged within $-0.1\leq K \leq 0.1$ for the H0L plane and $-0.1\leq H \leq 0.1$ for the 0KL plane. There are two twins with opposite tilts in the sample. Alignment was done with the dominant twin, which occupies an estimated 85-90\% of the crystal as determined from the ratios of various Bragg peak intensities. Since shearing of layers along the a-direction in the ac-plane corresponds to a displacement of Bragg peaks and other reciprocal-space features along L in the H0L plane, the Bragg peaks of the minority twin occupy positions of approximately (H,0,L-0.3H).  We accounted for the domain mosaic spread seen in the H0L plane in our data by convoluting the simulated intensities with a 2-dimensional Gaussian having widths in the radial and angular directions increasing linearly in the momentum transfer magnitude $|Q|$, with the mosaic spread estimated to correspond to that of the data. Since the simulated data were calculated for an orthorhombic supercell while the phase at 260 K is monoclinic, the 260 K simulated data were plotted for a monoclinic unit cell with $\beta\approx93.9^{\circ}$ by shifting intensities from (H,0,L) to (H,0,L-0.15H). The data plotted in Fig. 3(d) are averaged within $\pm 0.1$ r.l.u. in the H and K directions, subtracted from a background taken as the average of intensities along (1.8,0,L) and (2.2,0,L) which were each averaged within 0.05 and 0.1 r.l.u. in the H and K directions, respectively. To correct for misalignment during the temperature change, the simulated intensities were translated uniformly along L to match the data. The data in Fig. 3(e) were averaged similarly to Fig. 3 (d), but with the background taken as the average of (2.8, 0, L) and (3.2, 0, L) intensities. For the band structure calculations, we used density functional theory as implemented in the Vienna Ab initio Simulation Package (VASP). The details of the calculation are the same as those reported in Ref. \cite{cite29}.

Shown in Fig. 1(b) is a plot of the (HK0) scattering plane for data collected at 100 K in the T$_d$ phase. The Bragg reflections are the black dots while the red circles represent the calculated peak intensity. A similar plot is shown in Fig. 1(e) for data collected at 295 K in the 1T$\prime$ phase. The Rietveld refinement was performed on the structure factors extracted from Bragg reflections and the results are summarized in Table I. Ab initio calculation were performed on the refined structures of T$_d$ and 1T$\prime$ phases. With $k_z\sim0$, cuts near the Fermi surface are shown in Figs. 1(c) through 1(f) at $E\sim57 meV$ above the Fermi level, $E_F$. Weyl nodes, indicated by WP on the plots, are only observed in the T$_d$ phase (Fig. 1(c)), as previously reported, at the intersections of electron and hole pockets \cite{cite5,cite6,cite7}. In the centrosymmetric P2$_1$/m symmetry (Fig. 1(f)), Weyl nodes are absent, consistent with previous results as well \cite{cite5,cite6,cite7}. Thus the emergence of Weyl nodes upon cooling to the T$_d$ phase is tightly linked to the shifting layers, which is discussed next.

A single layer in MoTe$_2$ consists of the transition metal (Mo) atoms surrounded by the chalcogen (Te) atoms in either trigonal prismatic (2H) or octahedral (1T$\prime$ and T$_d$) local environments as shown in Fig. 1 (a) and (d). The refinement indicates that the 1T$\prime$ to T$_d$ transition that occurs between 260 and 240 K does not affect the local octahedral structure within the monolayer. Instead, the transition is driven by a relative shift of the layers along the ab-plane, changing the monoclinic unit cell to orthorhombic. The layer shift occurs between two high symmetry positions. In the 1T$\prime$ phase, the 2-fold screw rotation along the b-axis maps each layer to its next nearest neighboring layer. In the T$_d$ phase, the 2-fold screw rotation along the b-axis is broken by the shifting layers, while a new 2-fold screw rotation along the c-axis connecting adjacent layers is established.

Fig. 2 is a plot of intensity maps from elastic scattering in the H0L and 0KL planes. At 295 K, the crystal is in the 1T$\prime$ phase. At 260 K, the crystal is still in the 1T$\prime$ phase, but diffuse scattering streaks are observed along L in the H0L plane. By 240 K the crystal is mostly in the T$_d$ phase, but with some diffuse scattering streaks along L observed in the H0L plane that are less intense than at 260 K. By 100 K, the crystal has transformed into the T$_d$ phase to the point where no diffuse scattering intensity can be observed. The 0KL scattering planes shown on the right panels of Fig. 2 show no clear diffuse streaks along L at any temperature, in contrast to the diffuse scattering observed in the H0L plane. Diffuse streaks were also observed about the c-axis during the 1T$\prime$-T$_d$ transition by earlier X-ray diffraction measurements, but attributed to variations of the $\beta$-angle with temperature \cite{cite28}, which is different from what we discuss below.

To estimate the degree of stacking disorder during the structural phase transition, we compare the diffuse scattering intensity at 260 and 240 K to the results from a model of stacking disorder that we describe below. Both the 1T$\prime$ \cite{cite21} and T$_d$ \cite{cite23} phases can be constructed by stacking variants of a "base" layer, ignoring the slight differences in atomic coordinates within each layer between the T$_d$ and 1T$\prime$ phases. Each phase can be built from a sequence of stacking operations, "AAAAAA..." for T$_d$ and "ABABAB..." for 1T$\prime$ (Fig. 3(a)). Operation A reflects about the a-direction and translates by 0.5 lattice constants along the c-direction; the translation along the b-axis of operation A is 0.5 for T$_d$ and is different for the P2$_1$-symmetry 1T$\prime$, but this difference does not affect the diffuse scattering in the H0L plane. Operation B is the same as operation A but with an additional translation along the a-axis by $\pm0.15$ lattice constants, with the direction alternating from one layer to the next. The change in a-direction displacements results in the orthorhombic T$_d$ phase becoming monoclinic in 1T$\prime$.

For modeling the diffuse scattering, we start from the 1T$\prime$ stacking sequence ABABAB...., then randomly replace B-boundaries with A-boundaries with probability p. Although this model is crude, it allows us to verify that disordered A/B stacking does indeed explain the diffuse scattering reasonably well, and to estimate the amount of stacking defects at 260 and 240 K. A supercell is constructed with a large number of layers, and the intensity is taken as the square of the neutron scattering structure factor (with the Debye-Waller factor neglected for simplicity) \cite{cite31}, then convoluted with a Gaussian function to mimic resolution effects or changes in the mosaic spread. The diffuse scattering corresponds to intensities at fractional $L$ as defined in T$_{d}$-cell reciprocal lattice units. We note that only disorder for displacements along b-axis would contribute to diffuse scattering in the $0KL$ plane, which suggests that the lack of clear diffuse scattering streaks in the $0KL$ planes in Fig.\ \ref{fig:Slices} implies that there is little to no disorder along the $b$-direction. We also note that, since the additional $a$-axis displacements are multiples of $0.15\approx 1/6$, the contribution to the diffuse scattering along $(60L)$ is likely to be small. We indeed see little diffuse scattering along (60L), so the data appear to be consistent with our model. (Details on the relation between displacements along specific axes and diffuse scattering in specific planes can be found in the Supplement).

Simulated $H0L$-plane intensity maps are shown in Fig.\ \ref{fig:Model}(b) (260 K) and Fig.\ \ref{fig:Model}(c) (240 K). Both simulations were based on a 1000-layer supercell where a 1T$^{\prime }$\ crystal with two twins having volume fractions of 87\% and 13\% has B-boundaries replaced by A-boundaries with probability $p=0.3$ at 260 K and $p=0.8$ at 240 K. The model describes well the overall pattern of diffuse scattering in the $H0L$ plane; for example, similar diffuse scattering streaks are seen along $(20L)$, $(30L)$, and most of $(40L)$, while $(60L)$ and $(70L)$ are relatively clean. A closer look at the comparison between the data at 260 K and the model along (20L) is shown in Fig. 3 (d), again with calculations using a 1000-layer supercell and $p=0.3$. There is very good agreement between the model and data, with discrepancies likely due to either varying the distribution of domain orientation or due to inhomogeneities (different regions of the crystal having varying values of p).

For the 240 K data shown in Fig. 3 (e), we compare the intensity along (30L) to two models. Model \#1 is the model described above with $p=0.8$. Model \#2 is a similar model but starting from the T$_d$ phase stacking and replacing A-boundaries with B-boundaries with probability p=0.1. Both calculations used 4000 layers. These two models are motivated by different schemes of the 1T$\prime$ to T$_d$ transition. Model \#1 assumes B to A transitions are one-way and irreversible, whereas model \#2 is motivated by a transition scenario involving fluctuating layers which can shift back and forth but have an average probability of occupying at a certain position. Our data are more consistent with the first model than the second. First, we would expect the intensity upon (H, 0, L) to (H, 0, -L) to be symmetric in model \#2 but not in model \#1. (See Supplement for further details.) Although our data do not match perfectly with either model, intensities in the H0L plane appear to lack L-reflection symmetry, in particular intensities near (3, 0, -0.5) and (3, 0, 0.5) in Fig. 3 (e). Second, one would expect the distribution of twin domains to become more equal if A to B fluctuations were to occur with a significant probability, but our analysis suggests the estimated 87$\%$-13$\%$ distribution of twin domains at 295 K is most likely true at 260 K as well. To conclude, our layer stacking models can explain the diffuse scattering across the transition where the shift of the layers along the c-axis is coupled with Weyl node creation or annihilation.  It is this layer shift that breaks the 2-fold screw rotation along b-axis and establishes a new 2-fold screw rotation along c. This work may stimulate further studies to determine exactly at which angle and layer sequencing Weyl nodes are created or annihilated.

\section*{Acknowledgements}
This work has been supported by the Department of Energy, Grant number DE-FG02-01ER45927.A portion of this research used resources at the Spallation Neutron Source, a DOE Office of Science User Facility operated by the Oak Ridge National Laboratory.

\section*{Supplement}
\section{Obtaining Structure Factors from Diffraction Pattern}
The refinement of single crystal diffraction data requires obtaining structure factors from the Bragg reflections first. By indexing the Bragg peaks, an orientation matrix (a.k.a. UB matrix) is defined for the data at each temperature. The lattice constants are obtained from the UB matrices. For the 295 and 260 K data, the UB matrices are defined using only the reflections from one of the twin domains. The structure factors of each Bragg reflections are determined from the reduced and normalized data in order to determine the crystal structure. To accommodate the L-direction elongated peak shape, the intensities of the reflections are taken as integrals over a box of size $0.3 \times 0.3 \times 0.5$ in reciprocal units subtracting the average background in a shell of 0.08 reciprocal unit thickness. The refinement was performed on GSAS-I based on the observed structure factors, and the refined structures are listed in Table I.

\section{Mathematical details of structural transition}
As discussed in the main text, we compared the diffuse scattering intensity along $L$ at 240 and 260 K to a model where T$_{d}$-like layers are stacked on top of each other, with every other layer shifted by 0.15 lattice units along $a$ to form the 1T$^{\prime}$\ phase. A boundary between layers without a shift is labeled ``A'', and ``B''-boundaries are those with a relative shift along $a$. In our model, the initial structure is 1T$^{\prime}$, i.e., with ABABAB...\ stacking, but B is replaced by A with probability $p$. The intensity is given by the square of the structure factor, which is [31]: 

\begin{equation*}
F(\mathbf{G}) = \sum_{j} b_j e^{i \mathbf{G} \cdot \mathbf{d}_{j}},
\end{equation*}

where $j$ is the index for every atom in the supercell, $b_j$ is its neutron scattering length, $\mathbf{d}_j$ is its position, and $\mathbf{G}$ is a supercell reciprocal lattice wavevector (with fractional $L$-values in the units of the original T$_{d}$ unit cell). For simplicity, the Debye-Waller factor is neglected. We can now see that disordered displacements along the $a$-direction would only contribute to diffuse scattering in the $H0L$ plane, and disordered displacements along the $b$-direction would only contribute in the $0KL$ plane. The lack of clear diffuse scattering along $L$ in the $0KL$ plane and its presence in the $H0L$ plane suggests that the disordered displacements are primarily along the $a$-direction.

We can further clarify the situation by grouping atoms into layers: 

\begin{equation*}  \label{eq:StrFacLayers}
F(\mathbf{G}) = \sum_{j_3} F_{j_3} (\mathbf{G}) e^{i G_1 \delta_{j_3} a_1}e^{i G_3 j_3 a_3 / 2}.
\end{equation*}

Here, $F_{j_3} (\mathbf{G})$ is the structure factor sum over the atoms of each layer $j_3$, alternating between the two layer orientations; $a_1$ and $a_3$ are the T$_{d}$-cell lattice parameters along the $a$- and $c$-axes; and $G_1$ and $G_3$ are the components of $\mathbf{G}$ along these axes. The parameter $\delta_{j_3}$ encodes the disorder, representing the accumulated displacement in lattice units along the $a$-direction of the $j_3$th layer. For pure T$_{d}$, $\delta_{j_3} = 0$ for all $j_3$, and for pure 1T$^{\prime}$, for every other layer, $\delta_{j_3}$ changes by $\pm$0.15 (with the sign depending on the twin). For our model, $\delta_{j_3}$ can only have $0.15 n$ for some integer $n$. We can now explain the lack of diffuse scattering seen along $(60L)$; since $0.15 \approx 1/6$, $e^{i G_1 \delta_{j_3} a_1}$ becomes close to unity. In fact, the lack of diffuse scattering along $(60L)$ at either 240 or 260 K eliminates any model where disordered displacements along the $a$-direction do not approximately follow $\delta_{j_3} = n/6$.

In addition to our main model (labeled  \textquotedblleft model \#1\textquotedblright in Fig. 3(e)), we made comparisons with a related model, \textquotedblleft model \#2\textquotedblright , which differs in that the initial structure is AAAA...\ (the T$_{d}$ phase) and that there is a $p$-probability of converting A-boundaries to B-boundaries. If B$\rightarrow $A transitions during the 1T$^{\prime }\rightarrow $T$_{d}$ transition are assumed to be one-way and irreversible, then model \#1 would apply, whereas model \#2 would make more sense if the layers were thought to be fluctuating back and forth, on average being in a certain state with a certain probability. One important difference between these models is the expected symmetry of the diffuse scattering for $L\rightarrow-L$. In each case, the total structure factor can be split into components for the even- and odd-numbered layers (in order of stacking). For model \#2 in an infinite-layer supercell, these two components transform into each other (up to a translation) upon reflection in the $a$-direction, implying that the scattering intensity obeys a $H\rightarrow -H$ symmetry, which implies a $L\rightarrow -L$ symmetry in the $H0L$ plane by Friedel's law. For model \#1, the even- and odd-layer components do not transform into each other upon $a$-direction reflection since only one of those sets of layers has disordered displacements; thus, for model \#1 we expect asymmetry for $(H,0,L)\rightarrow (H,0,-L)$, which is consistent with our data.

\bibliographystyle{unsrt}

\begin{figure}[h]
\begin{center}
\includegraphics[width=8.6cm]{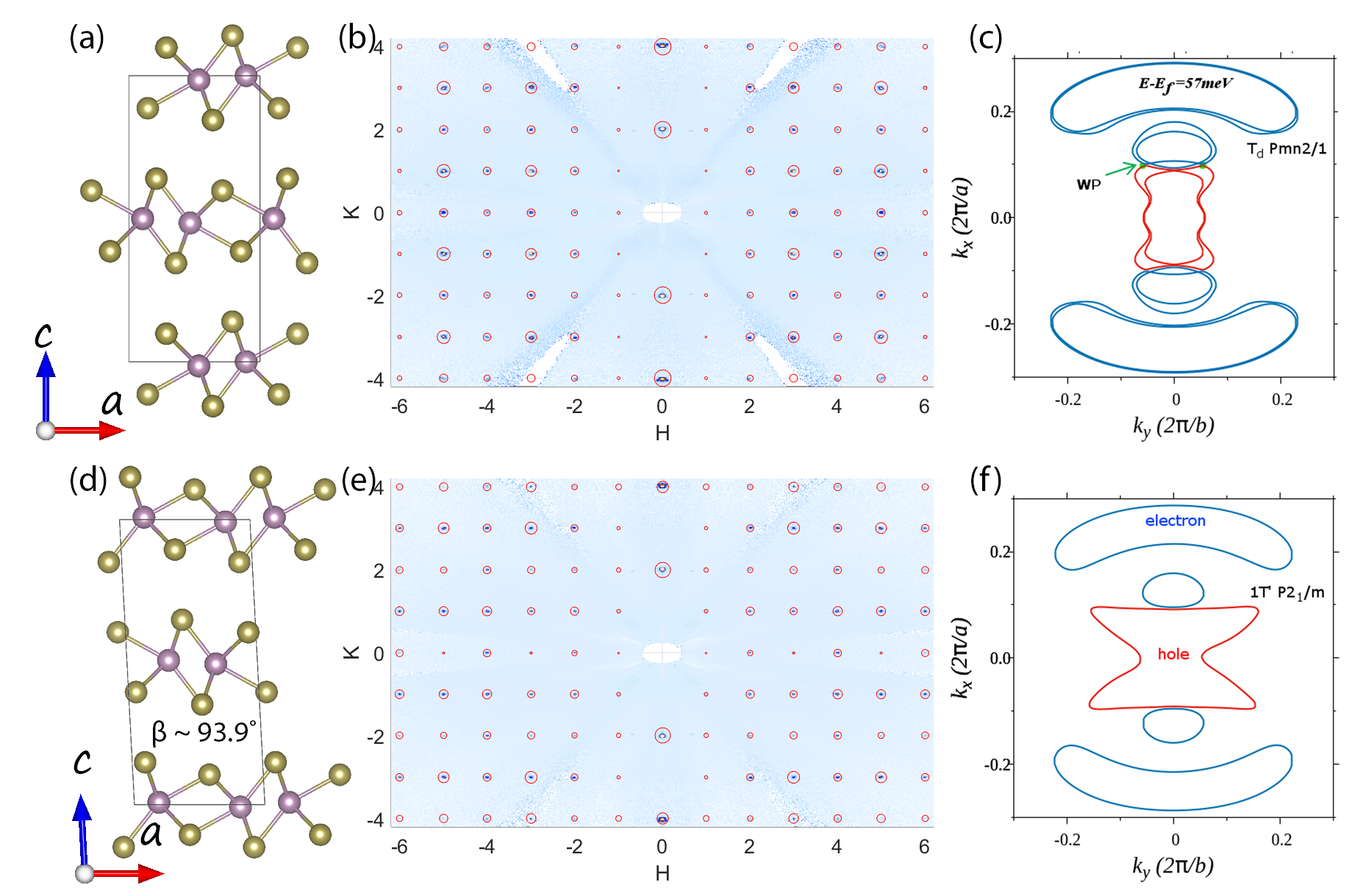}
\end{center}
\caption{Shown in (a) and (d) are the refined crystal structures of the T$_d$ and 1T$\prime$ phases of MoTe$_2$. (b) and (e) are neutron scattering intensity maps of the HK0 planes at 100 K and 295 K. (c) and (f) are plots in the $k_x$-$k_y$ plane of the Fermi surface with $k_z\sim0$ from electronic band structure calculations using the refined parameters of Pnm2$_1$ and P2$_1$/m structures.}
\label{fig:P1Sym}
\end{figure}

\begin{figure}[h]
\begin{center}
\includegraphics[width=8.6cm]{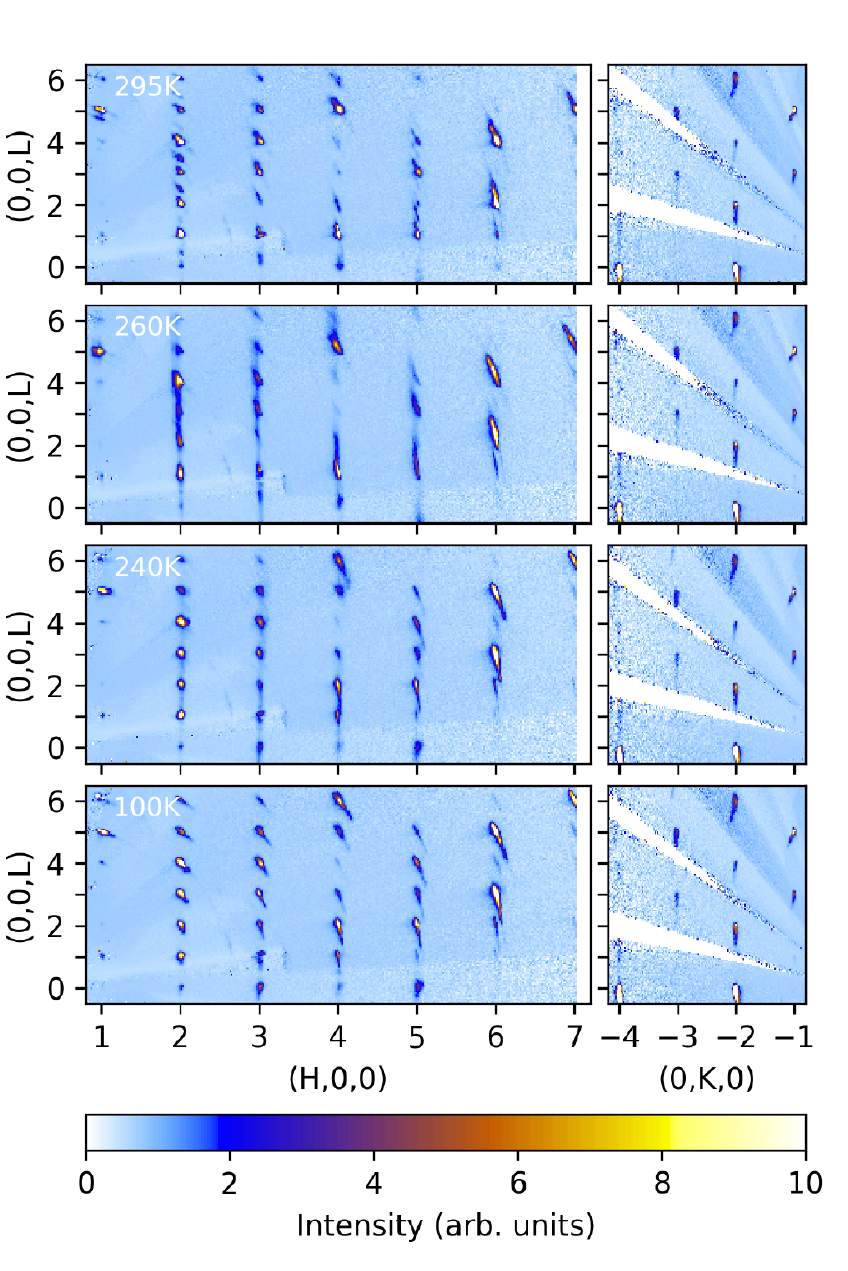}
\end{center}
\caption{Neutron scattering intensity maps of the H0L and 0KL planes. Diffuse scattering streaks appear along L in the H0L plane only, upon cooling from 295 to 240 K. By comparison, in the 0KL plane, no diffuse streaks are observed. }
\label{fig:Slices}
\end{figure}

\begin{figure}[h]
\begin{center}
\includegraphics[width=8.6cm]{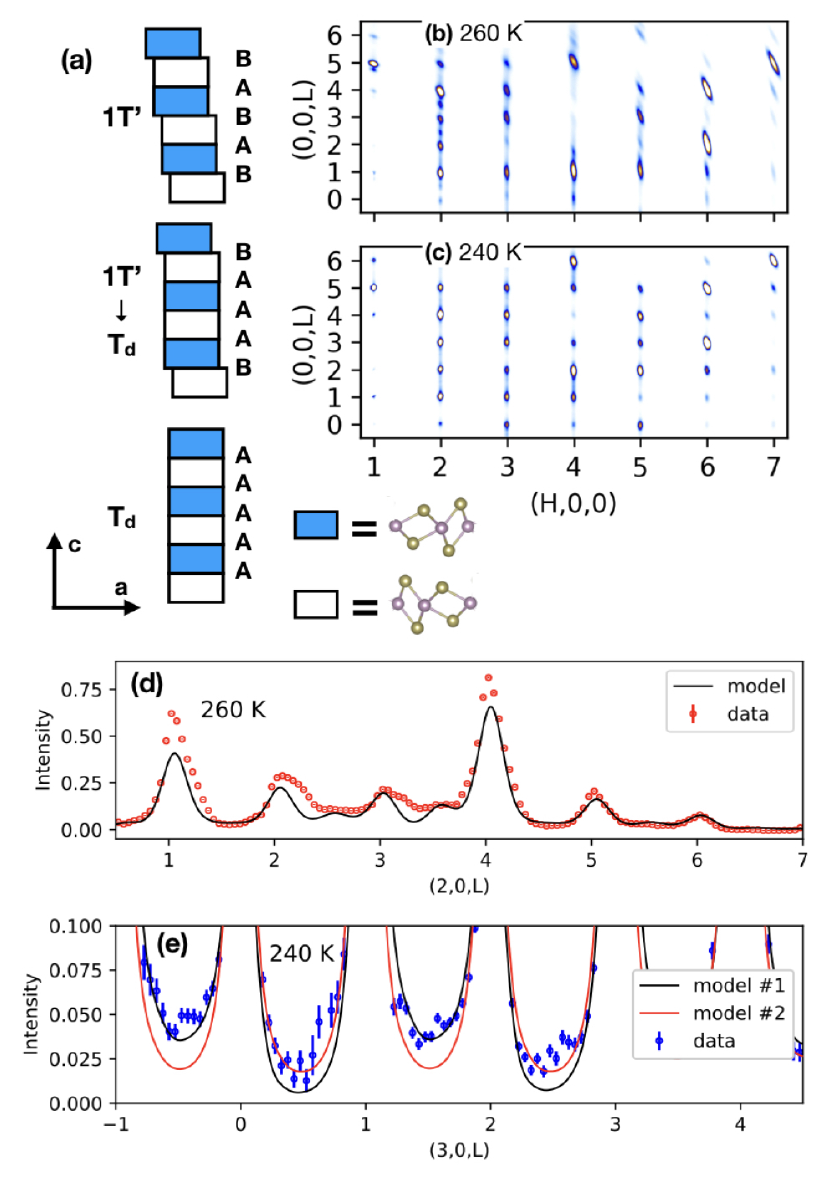}
\end{center}
\caption{(a) A schematic diagram describing the model used to explain the diffuse scattering. (b, c) Simulated H0L intensity maps for 260 K in (b) and 240 K in (c). (d) Comparison of model and data along (20L) at 260 K. (e) Comparison of the diffuse scattering along (30L) at 240 K and the results of the same model as for (c) (\# 1), with a second model (\# 2).}
\label{fig:Model}
\end{figure}

\begin{table}[b]
\caption{Refined atom positions of MoTe$_2$ at 295 K, 260 K, 240 K and 100 K using P2$_1$/m (295 K and 260 K)  and Pnm$2_1$ (240 K and 100 K) space group. The lattice constants are $a=6.33$ \AA, $b=3.48$ \AA, $c=13.82$ \AA, $\beta=93.8 ^{\circ}$ (295 K), and $\beta=93.7 ^{\circ}$ (260K).}
\label{tab:table1}
\begin{ruledtabular}
		\begin{tabular}{ccccc}
			&	 295 K	& 260 K & 240 K & 100 K\\
			\hline
Mo1 $x$		&  0.188(1)	 &0.191(1) &0.8975(7)&0.8944(6) \\
\hphantom{Mo1} $z$		 &0.0078(5) &0.0081(6)& 0.0001(4) &0.9996(3)\\
Mo2 $x$	    & 0.319(1)	& 0.318(1)&0.5316(7)&0.5288(5)\\
\hphantom{Mo1} $z$	    & 0.5066(5) &0.5062(7) &0.9857(3)&0.9859\\
Te1 $x$		& 0.582(1)	&0.579(2)&0.287(1)&0.2838(8)\\
\hphantom{Mo1} $z$		& 0.1050(6)	&0.1052(8)&0.0979(5)&0.0979(4) \\
Te2	$x$	    &0.104(1)	&0.107(2)&0.7926(9)&0.7893(7)\\
\hphantom{Mo1} $z$	    &0.1505(6)	&0.1513(7)&0.1416(5)&0.1410(3)\\
Te3	$x$		& 0.556(1)	&0.558(2)&0.3589(9)&0.3610(7)\\
\hphantom{Mo1} $z$		& 0.3512(6)	&0.3511(7)&0.3437(4)&0.3444(3)\\
Te4 $x$		& 0.053(1)	&0.054(2)&0.856(1)&0.8592(7)\\
\hphantom{Mo1} $z$		& 0.3955(6)	&0.3963(7)&0.3870(5)&0.3873(3)\\
			\hline
$\chi^2$	&22.28	&28.72&6.72&4.94\\
		\end{tabular}
	\end{ruledtabular}
\end{table}

\end{document}